\documentclass[12pt,a4paper]{article}

\RequirePackage{ifpdf} 
\usepackage{amsmath} 
\usepackage{mathtools}
\usepackage{youngtab}

\usepackage{jheppub}
\usepackage{pstricks}
\usepackage[final]{pdfpages} 
\usepackage{ifpdf} 
\usepackage{slashed}
\usepackage{hyperref}

\usepackage{color} 
\usepackage{graphics}

\usepackage{etoolbox} 
\usepackage{fixmath}

\usepackage{caption} 
\usepackage{subcaption} 
\usepackage{amsfonts}

\usepackage{multirow}
\usepackage{epstopdf}

\usepackage{relsize}
\usepackage{float}

\usepackage{tikz}
\usetikzlibrary{positioning,arrows}
\usetikzlibrary{decorations.pathmorphing}
\usetikzlibrary{decorations.markings}
\usetikzlibrary{shapes.geometric}
\usepackage{endnotes}
\title{Bi-partite vertex model and multi-colored link invariants}

\author{Saswati Dhara$^{a}$, Romesh K. Kaul$^{b}$, P. Ramadevi$^{a}$,  and Vivek Kumar Singh$^{a}$} 

\affiliation{$^a$ Department of Physics, Indian Institute of Technology Bombay,Mumbai 400076, India  \\ 
 $^b$ The Institute of Mathematical Sciences, Chennai 600113, India}

\emailAdd{saswati123@phy.iitb.ac.in}
\emailAdd{kaul@imsc.res.in}
\emailAdd{ramadevi@phy.iitb.ac.in}
\emailAdd{viveksingh@phy.iitb.ac.in}

\abstract{Construction of representations of braid group generators from $N$-state vertex models provide an elegant route to study knot and link invariants. Using such a  braid group representation, an algebraic formula for the link invariants was put forth when the same spin $(N-1)/2$ are placed on all the component knots. In this paper, we generalise the procedure to deduce representations of braiding generators from  bi-partite vertex models. Such a representation allows the study of multi-colored link invariants where the component knots carry different spins. We propose a multi-colored link invariant formula in terms of braiding generators derived from $R$ matrices of  bi-partite vertex models.
}
\begin{document}



\allowdisplaybreaks[4]
\unitlength1cm
\maketitle
\flushbottom

\let\footnote=\endnote
\renewcommand*{\thefootnote}{\fnsymbol{footnote}}

\section{Introduction}
Both mathematicians and physicists  have attempted efficient methods of obtaining a polynomial form of knot and link invariants. As knots and links can be obtained from a closure of braid word, the knot and link  invariants can be derived from the representation theory of braid groups.  Interestingly, there are diverse approaches attempting different braid group representations. 

The pioneering work of Witten\cite{witten} on $SU(2)$ Chern-Simons theory is one such approach where the Wilson loop  expectation value $\langle W_R(\mathcal {K})\rangle$  reproduces  the  Jones' polynomial\cite{jones} for representation $R={\tiny \yng(1)}$ placed on the knot  $\mathcal {K}$. The main ingredient in this approach is the relation between three-dimensional $SU(2)$ Chern-Simons theory and two-dimensional $SU(2)_k$ Wess-Zumino Witten (WZW) model where $k$ is the Chern-Simons coupling constant which determines the level of WZW model. The monodromy  matrices along with fusion matrices(duality matrices) of the WZW model provide representations for the braid group.  In fact, this procedure can be generalised for other gauge groups $\mathcal G$  but the knowledge of duality matrices for arbitrary representations $R$ are not known. 

The primary fields of the WZW model are in one to one correspondence with the finite dimensional unitary representations of  quantum groups $U_q(\mathcal G)$  where the deformation parameter $q$ is chosen to be root of unity. Further, the operator product expansion of these primary fields resembles tensor products of representations  in the context of quantum group. The level $k$ is related to $q$. For instance, $q=\exp({2\pi i/k+2})$  for $SU(2)$ group \cite{kir}. The universal  $\mathcal R$ matrices  constructed using the generators $J^{\pm}, J_z$ of $U_q(SU(2))$ indeed obey defining relations of braid groups. Hence, these universal $\mathcal R$ matrices provide representations for braid groups.

Exactly solvable statistical mechanical models\cite{baxter} appears completely different approach towards construction of braid group representation. $N$-state vertex models are one such statistical mechanical model with Boltzmann weights $(R^{j,j})_{m_1 m_2}^{n_1,n_2}(u)$   associated with every vertex (see Fig.1(a)), on a square lattice, depending on the states $m_1,m_2,n_1,n_2 \in \{-j,-j+1,\ldots j\}$  placed on four edges intersecting the vertex where 
the spin  $j=(N-1)/2$. The Yang-Baxter equations are obeyed by spectral parameter $u$  dependent Boltzmann weights $(R^{j,j})_{m_1 m_2}^{n_1,n_2}(u)$ of these vertex models. In fact, the Yang-Baxter equation in the limit of $u \rightarrow \infty$ can  be reduced to defining relations of braid group by applying permutation operator $\hat P$  on the Boltzmann weights. That is., the braiding generators are proportional to  $ \hat P[(R^{j,j})_{m_1 m_2}^{n_1,n_2}(u\rightarrow \infty)]$. 

The algebraic expression for the knot polynomial using these braiding matrices have been studied in Refs.\cite{akutsu,akutsu1,akutsu2,deguchi,deguchi1,deguchi2} for spin $j=1/2, j=1,j=3/2$ which are also known in the literature as $6$-vertex, $19$-vertex and $44$- vertex models respectively. The numbers $6,19,44$ indicate the count of non-zero Boltzmann weights for the corresponding N-state vertex model. The polynomial form can be computed for any braid word $A$ using the algebraic expression $\alpha_{\small{\underbrace{j,j,\ldots j}_n}}(A)$ (\ref{formu4}) as discussed  in Refs.\cite{akutsu,akutsu1,akutsu2,deguchi,deguchi1,deguchi2}. Here spin $j$ states are placed on all the strands of the braid and $n$ denotes the number of component knots of the link obtained from closure of braid $A$.

It is also important to relate these braiding matrices with  the monodromy matrices in WZW models.  Using the quantum deformed Clebsch-Gordan coefficients $q-CG$, these braiding matrices $\hat P [(R^{j,j})_{m_1 m_2}^{n_1,n_2}(u\rightarrow \infty)]$ can be diagonalised\cite{kaul1} whose diagonal elements $\lambda_J(j,j)$ are the eigenvalues of monodromy matrices in the WZW model.  Extending this diagonalization procedure to the  known Boltzmann weights of the $6$-vertex, $19$-vertex and $44$-vertex models, the  spectral parameter dependent diagonal matrix elements $\lambda_J(j,j;u)$ can be obtained. From these examples of vertex models Boltzmann weights and their diagonalization, it was straightforward to conjecture spectral parameter dependent diagonal matrix elements for spin $j>3/2$. Interestingly, the procedure can be reversed resulting  in deducing Boltzmann weights for new vertex models where the edges carry states of spin $j>3/2$\cite{kaul1}.  

Thus, these  vertex models provide us new representations of braiding matrices which is useful to construct new knot invariants using the algebraic expression $\alpha_{\small{\underbrace{j,j,\ldots j}_n}}(A)$\cite{akutsu,akutsu1,akutsu2,deguchi,deguchi1,deguchi2}. We would like to emphasize that obtaining knot and link polynomials using such an algebraic expression is definitely an efficient approach as they involve only multiplication of matrices corresponding to any arbitrary braid word $A$.

Just like the conventional vertex model approach enables efficient computation of  knot and link polynomials $\alpha_{\small {\underbrace{j,j\ldots j}_n}}(A)$, we  wanted to attempt a modified algebraic expression for link invariant  $\alpha_{\small {j_1,j_2\ldots j_n}}(A)$ in terms of matrix representations of braiding matrices  where $j_1,j_2, \ldots$ are the spin states placed on different component knots of a  link. We will refer to these link invariants as {\bf multi-colored link invariants.} 

\begin{figure}
\centering{\includegraphics[scale=0.2]{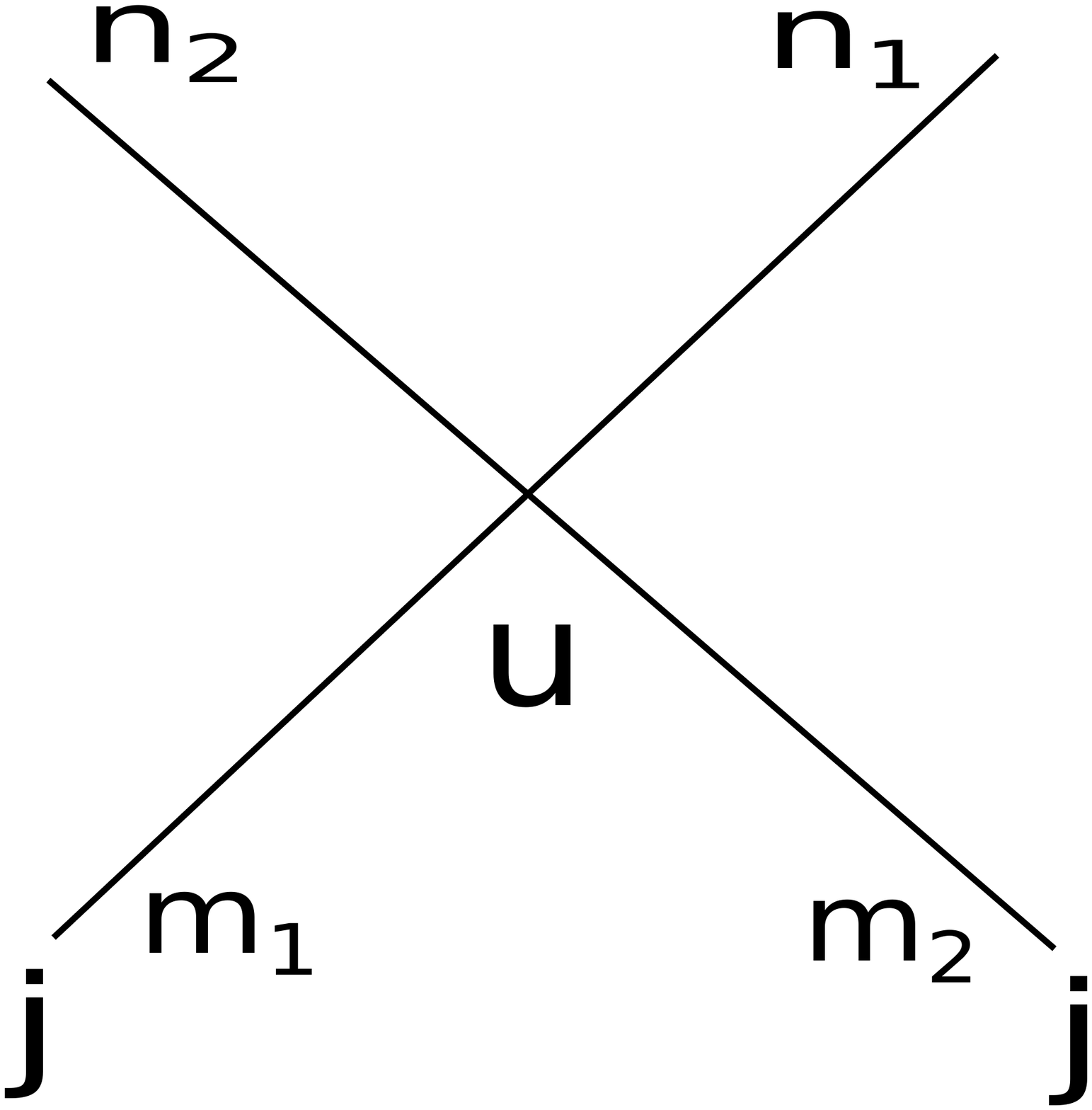}~~~~~~~~~~~~~~~~~~~~~~~~~\includegraphics[scale=0.2]{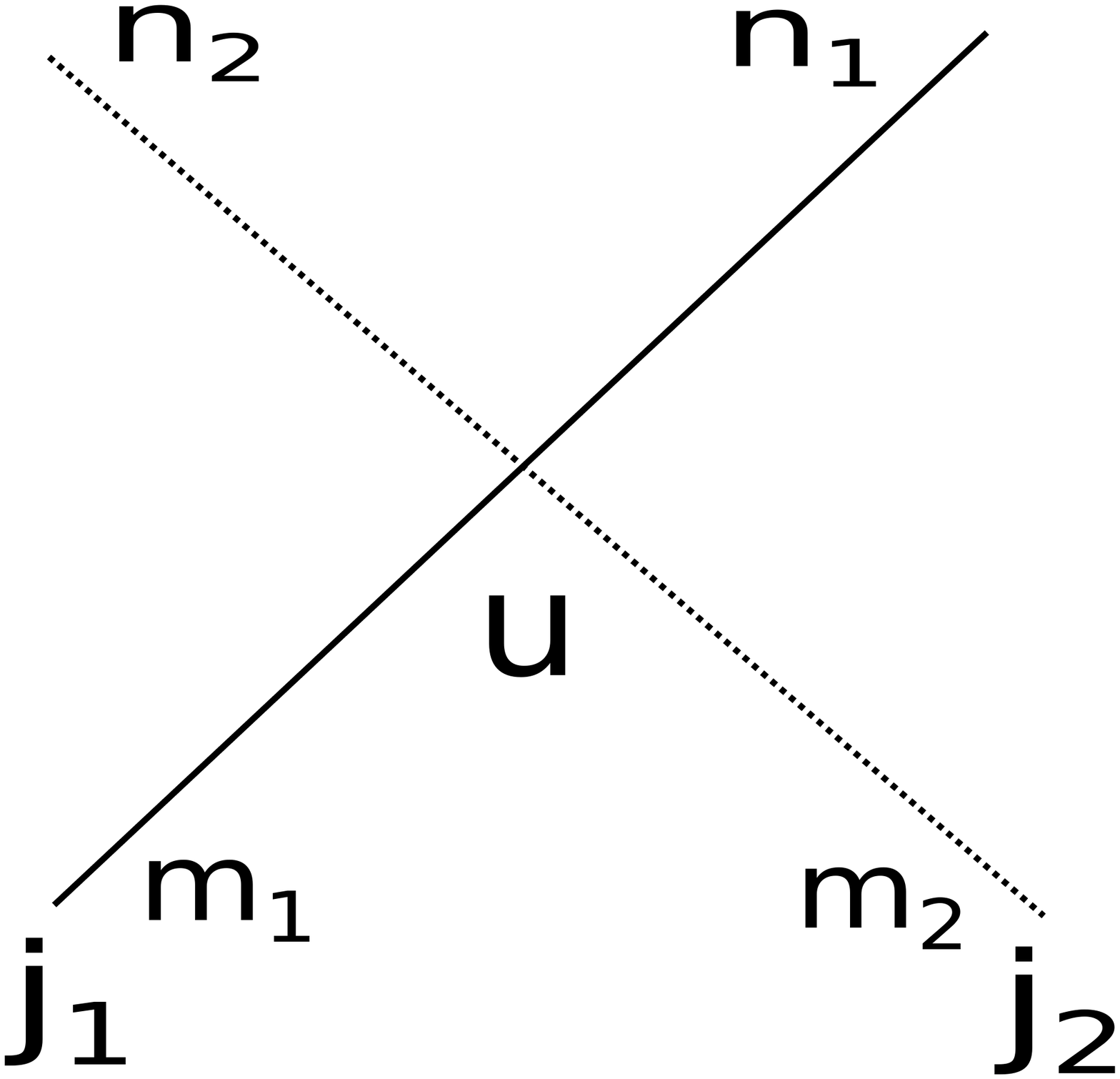}}
\caption{(a)Vertex model~~~~~~~~~~~~~~~~~~~~~~~~~~~ (b)Bi-partite Vertex model\\
$~~~~~~~~~~~~~~~~~~~(R^{j,j})_{m_1,m_2}^{n_1,n_2}(u)$~~~~~~~~~~~~~~~~~~~~~~~~~~~~~~~~~~~~~~$(R^{j_1,j_2})_{m_1,m_2}^{n_1,n_2}(u)$}
\label{Rm}
\end{figure}

The braiding generators must be derivable from $(R^{j_1,j_2})_{m_1 m_2}^{n_1n_2}(u)$-matrices of new vertex models whose vertices are of type as shown in fig.~\ref{Rm}(b). We call such vertex models as {\bf bipartite vertex models} where  two of the four edges carry states $m_1,n_1 \in {j_1,j_1-1, \ldots -j_1}$ and the other two edges  carry   $m_2,n_2 \in {j_2,j_2-1, \ldots -j_2}$.

The procedure for obtaining vertex model Boltzmann weights $(R^{j,j})_{m_1,m_2}^{n_1,n_2}(u)$ from spectral parameter dependent diagonal braiding matrix elements  $\lambda_J(j,j;u)$ can be generalized to deriving  bi-partite vertex models Boltzmann weights  $(R^{j_1, j_2})_{m_1,m_2}^{n_1,n_2}(u)$ where two adjacent edges carry different spins $j_1,j_2$ as illustrated in Fig.1(b). It is pertinent to mention that there will be groupoid relations\cite{kaul1994} for braid generators  $\hat P (R^{j_1, j_2})_{m_1 m_2}^{n_1,n_2}(u \rightarrow \infty))$ obtained from Boltzmann weights in the $u \rightarrow \infty$ limit with the application of a suitable permutation operator $\hat P$.   
Our main focus in this paper is to  construct representations of braids from $(R^{j_1,j_2})_{m_1 m_2}^{n_1,n_2}(u)$  which obeys groupoid properties\cite{kaul1994} due to different spin $j$ states on the strands. 
This methodology of deducing matrix form for braiding generators from bi-partite vertex model Boltzmann weights lead to efficiently compute multi-colored link polynomials $\alpha_{\small {j_1,j_2\ldots j_n}}(A)$ using our modified formula for links from the closure of arbitrary braid word $A$. 
 
Plan of the paper is as follows.  In sec.\ref{vertexm},  we will review the construction of  braiding matrices from  $R$-matrices  of conventional vertex models for same spin case and the derivation of knot and link invariants. In sec.\ref{bipartv}, we will generalise the procedure to determine new representation of braiding generators from $R(u)$-matrices associated with bi-partite vertex models and propose an algebraic formula for multi-colored link invariants. We will summarise and suggest open problems in the concluding section.

\section{Vertex models \& R-matrix}\label{vertexm}
In this section, we will briefly review vertex model approach of constructing representations of braid group generators leading to evaluating knot or link invariants.
\subsection{Vertex model}
$N$-vertex models are two dimensional statistical mechanical model with states of same spin $j$ placed on the four edges intersecting every vertex as  shown in fig.~\ref{Rm}(a). These statistical mechanical models are exactly solvable provided the spectral parameter $u$ dependent Boltzmann weights $(R^{j,j})_{m_1,m_2}^{n_1,n_2}(u)$ satisfies the following Yang-Baxter equation:
\begin{eqnarray}
\sum\limits_{m'_1,m'_2,m'_3}& &(R^{j,j})_{m_1,m_2}^{m_1',m_2'}(u)(R^{j,j})_{m'_1,m_3}^{m_1'',m_3'}(u+v)(R^{j,j})_{m_2',m_3'}^{m_2'',m_3''}(v)\nonumber\\  &=&\sum\limits_{m'_1,m'_2,m'_3}(R^{j,j})_{m_2,m_3}^{m_2',m_3'}(v)(R^{j,j})_{m_1,m_3'}^{m_1',m_3''}(u+v)(R^{j,j})_{m_1',m_2'}^{m_1'',m_2''}(u)~.
\label{ybe}
\end{eqnarray}
The parametrized form of these $R$-matrices dependent on the spectral parameter $u$ and another parameter $q=e^{2\mu}$ are given in \cite{baxter,akutsu} for $6,19,44$ vertex models. In the limit $u\rightarrow \infty$, the above equation involving $R$-matrix elements multiplied by a permutation operator $\hat{P}$ (upto an overall normalisation) will resemble defining relation of a braid group 
$\mathcal B_r$ where the generators $b_i$'s ($i=1,2,\ldots r$) have the following representation:
\begin{eqnarray}
b_i[j,j]&=&\underbrace{ \mathbb I_1 \times \mathbb I_2 \times \ldots \mathbb I\times}_{i-1}
 (\hat{R}^{j,j})_{m_1,m_2}^{n_1,n_2}\times \mathbb I_{i+2} \ldots \nonumber\\
b_i[j,j] ^{-1}&=&\underbrace{ \mathbb I_1 \times \mathbb I_2 \times \ldots \mathbb I\times}_{i-1} (\hat (\hat{R}^{j,j})_{m_1,m_2}^{n_1,n_2})^{-1}\times \mathbb I_{i+2} \ldots ~.\label{condn}
\end{eqnarray}
Here 
\begin{equation}
(\hat{R}^{j,j})_{m_1,m_2}^{n_1,n_2}={1 \over \mathcal N} \hat{P}(R^{j,j})_{m_1,m_2}^{n_1,n_2}(u\rightarrow \infty),
\label{formu3}
\end{equation}
where the normalisation factor   $\mathcal N=(R^{j,j})_{j,j}^{j,j}(u\rightarrow \infty)$  ensures that all the matrix elements are finite in the $u\rightarrow \infty$ limit. Thus we have new braid group representations from vertex model $R$-matrices leading us to new link invariants.

 The following algebraic formula\cite{akutsu} defines {\bf invariant} of any $n$-component link, obtained from the closure of braid word $A\in \mathcal B_r$,  with same spin $j$ on their component knots: 
 \begin{equation}
\alpha_{\small{\underbrace{j,j,\ldots j}_n}}(A)= (\tau_j \bar{\tau}_j)^{-n/2} \left(\frac{\bar{\tau}_j}{\tau_j}\right)^{(e/2)}Tr[HA],
\label{formu4}
\end{equation}
where $e$ is the exponent sum of the $b_i$'s  appearing in the braid word $A$,
\begin{eqnarray}
H &=& \underbrace{h_{j}\otimes h_{j}\ldots h_j}_r ~{\rm where}\nonumber\\
h_j&=&\frac{1}{1+q+\ldots+q^{2j}}~\rm Diag [1,q,\dots,q^{2j}]~,
\label{hmat}
\end{eqnarray}
and $\tau_j$ and $\bar{\tau}_j$ are 
\begin{equation}
\tau_j=\frac{1}{1+q+\ldots+q^{2j}}~;~\bar{\tau}_j=\frac{q^{2j}}{1+q+ \ldots+q^{2j}}~.
\label{tau}
\end{equation} 
The above invariant $\alpha_{j,j,\ldots j}(A)$, in variable $q$,  remains unchanged if we perform Markov moves on braids. Further,  we work with the following unknot invariant: $\alpha_j( b_1)=\sum_{i=-j}^j q^i$. These invariants are known in the knot theory literature as unnormalised link invariants.

In the following subsection, we briefly review the link invariant computation using the braiding matrices derived from the simplest $6$-vertex model  $R$-matrices. That is.,  the edges in fig.~\ref{Rm}(a) carry $j=1/2$.
\subsection{6-vertex models}\label{6v}
The simplest vertex model is the  $6$-vertex model where states of  $j=\frac{1}{2}$ are placed on the four edges intersecting every  vertex. So the Boltzmann weights $(R^{\frac{1}{2},\frac{1}{2}})_{m_1,m_2}^{n_1,n_2}(u)$ associated with every vertex are nonzero if and only if
$m_1+m_2=n_1+n_2$ where $m_1,m_2,n_1,n_2 \in \{-1/2,1/2\}$. This condition allows six non-zero Boltzmann weights which is kept track by calling the model as $6$-vertex model. 
In matrix form, the elements are :
$$ (R^{\frac{1}{2},\frac{1}{2}})_{m_1,m_2}^{n_1,n_2}(u)=\left(
\begin{array}{c|ccccccc}
m_1,m_2\backslash n_1,n_2\rightarrow          & \uparrow\uparrow & \uparrow\downarrow &\downarrow\uparrow & \downarrow\downarrow   \\
\hline
\uparrow\uparrow & \sinh(\mu-u) & 0 & 0 & 0   \\
\uparrow\downarrow            & 0 & -\sinh{u} & e^{u}\sinh{\mu} & 0   \\ 
\downarrow\uparrow         & 0 & e^{-u}\sinh{\mu} & -\sinh{u} & 0 \\
\downarrow\downarrow     & 0 & 0 & 0 & \sinh(\mu-u)  \\
\end{array}
\right).$$
In order to construct the braid generators $b_i$, we take the limit  $u\rightarrow\infty$ on the above matrix elements and replace $e^{2 \mu}$ by variable $q$. Further we choose a suitable normalisation such that the matrix elements are finite in this limit $u\rightarrow \infty$ as shown below: 
$$ {(R^{\frac{1}{2},\frac{1}{2}})_{m_1,m_2}^{n_1,n_2}(u\rightarrow \infty) \over(R^{\frac{1}{2},\frac{1}{2}})_{\uparrow,\uparrow}^{\uparrow,\uparrow}(u\rightarrow \infty)}
=\left(
\begin{array}{c|ccccccc}
m_1,m_2\backslash n_1,n_2\rightarrow          & \uparrow\uparrow & \uparrow\downarrow &\downarrow\uparrow & \downarrow\downarrow   \\
\hline
\uparrow\uparrow & 1 & 0 & 0 & 0   \\
\uparrow\downarrow            & 0 &  q^{1/2}& 1-q & 0   \\ 
\downarrow\uparrow         & 0 & 0 & q^{1/2}  & 0 \\
\downarrow\downarrow     & 0 & 0 & 0 & 1  \\
\end{array}
\right).$$
Using the following permutation matrix 
$$ \hat{P}^{1/2,1/2}=\left(
\begin{array}{cccc}
 1 & 0 & 0 & 0   \\
 0 & 0& 1 & 0   \\ 
 0 & 1& 0  & 0 \\
 0 & 0 & 0 & 1  \\
\end{array}
\right),$$
the elements of  $(\hat{R}^{\frac{1}{2},\frac{1}{2}})_{m_1,m_2}^{n_1,n_2}$ (\ref{formu3}) turn out to be
 $$ (\hat{R}^{\frac{1}{2},\frac{1}{2}})_{m_1,m_2}^{n_1,n_2}=\left(
\begin{array}{ccccccc}

 1 & 0 & 0 & 0   \\
 0 & 0 & q^{1/2} & 0   \\ 
 0 & q^{1/2}& 1-q  & 0 \\
 0 & 0 & 0 & 1  \\
\end{array}
\right).$$
Hence, we can determine the matrix form of the braid generators $b_i[1/2,1/2]$ (\ref{condn}) using the $\hat R$ matrix. We can work out the invariants (\ref{formu4}) for some knots and links. 
In Table.\ref{same}, we have listed the knots and links with their braid word and invariants in 
variable $q$.
\begin{table}
\begin{tabular}{|c|c|c|c|}
\hline 
Knot/Link & Braidword & e & Polynomial \\ 
\hline 
Trefoil $3_1$ & $b_1^3 \in \mathcal B_2$ &3& $-q^{1/2} (1+q) \left(-1-q^2+q^3\right)$ \\ 
\hline 
HopfLink & $b_1^2 \in \mathcal B_2$ &2& $q^{1/2} \left(1+q^2\right)$ \\ 
\hline 
Figure eight $4_1$  &$ b_1b_2^{-1}b_1b_2^{-1}\in \mathcal B_3$ &0& $q^{-5/2}(1+q^5)$ \\ 
\hline 
L7a3  &$ b_1b_2^{-1}b_1^3b_2^{-1}b_1\in \mathcal B_3$ &3& $q^{-1}+2 q+q^2+q^3-q^5+q^6-q^7$ \\ 
\hline
Whitehead  &$b_1.b_2^{-1}.b_1.b_2^{-2}\in \mathcal B_3$ & -1& $q^{-4}(-1+q+q^2+q^3+q^4+q^6)$ \\ 
\hline
Boromean &$b_1^{-1}b_2b_1^{-1}b_2b_1^{-1}b_2\in \mathcal B_3$&0 & $q^{-7/2}(-1+2 q+q^2+2 q^3+2 q^4+q^5+2 q^6-q^7)$ \\ 
\hline
\end{tabular} .
\caption{}{\label{same}}
\end{table}
Recall that there is only one braid generator $b_1$ for all  braid words $A\in \mathcal B_2$ whose matrix form will be $4 \times 4 $ matrix. That is.,
$$b_1=(R^{\frac{1}{2},\frac{1}{2}})_{m_1,m_2}^{n_1,n_2}$$
We have worked out  the invariants in eqn.(\ref{formu4}) for 
unknot, trefoil and Hopf links  using braid words $A=b_1,b_1^3 ~{\rm and}~ b_1^{2}$ respectively (see Table.1)

For knots and links obtained from closure of  braid words $A \in \mathcal B_3$, there are two braiding generators $b_1,b_2$ which are  $8\times 8$ matrices:
$$b_1=(R^{\frac{1}{2},\frac{1}{2}})_{m_1,m_2}^{n_1,n_2}\times \mathbb{I},$$ 
$$b_2=\mathbb{I}\times(R^{\frac{1}{2},\frac{1}{2}})_{m_1,m_2}^{n_1,n_2}. $$ 
For example, figure eight $4_1$ knot whose braid word is  $A=b_1^{-1}b_2b_1^{-1}b_2$.
We must remember that such a braid word action on a $3$-strand braid implies  
the following order of matrix operation on an initial state $\vert j,m_1;j,m_2;j m_3\rangle$:
\begin{equation}
A\vert {\rm 3-strand} \rangle \equiv b_2 \left[ b_1^{-1}\{b_2 \left(b_1^{-1} \vert j,m_1;j,m_2;j,m_3\rangle\right)\}\right]~.\label{ordbraid}
\end{equation}
The method can be generalised for any  braid word $A\in \mathcal B_n$  leading us to evaluate polynomial invariants (\ref{formu4}). Further these polynomials match with the Jones' polynomials upto unknot normalisation.  As the approach involves only multiplication of matrices,  this method is highly efficient in obtaining  polynomial invariants for any knot or link from vertex models whose $R$-matrix elements are known. 

In the literature, colored Jones' polynomials correspond to placing higher spins $j\geq 1$ on the component knots. Interestingly, these polynomials  for $j=1,3/2$  agree with the link invariant $\alpha_{j,j,\ldots j}(A)$ in eqn.(\ref{formu4}) where the matrix representation of the braid generators $b_i$'s are derived from Boltzmann weights of the $19-$vertex and $44$- vertex models. 

The  braiding generators $b_i$'s derived from $(R^{j,j})_{m_1,m_2}^{n_1,n_2}$-matrix of vertex models  as well as  from  eigenvalues ($\lambda$)  of the monodromy matrices in $SU(2)_k$ Wess-Zumino conformal field theory suggested a compact elegant relationship\cite{kaul1}:
\begin{eqnarray}
(\hat{R}^{j,j})_{m_1,m_2}^{n_1',n_2'}&=&\frac{1}{ \mathcal N }\hat{P}^{j,j}(R^{j,j})_{m_1,m_2}^{n_1,n_2}(u\rightarrow \infty)\nonumber\\&=&\frac{1}{ \mathcal N }\hat{P}^{j,j}\sum_{J \in  j \otimes j}
\Big \{\begin{matrix}
j & j & J\\
m_2 & m_1 & M 
\end{matrix}\Big\} \lambda_J(j,j)
\Big\{\begin{matrix} 
j & j & J\\
n_1 & n_2 & M 
\end{matrix}\Big \},
\label{rmat1}
\end{eqnarray}
where $M=m_1+m_2=n_1+n_2$ and the terms in parenthesis $\Big\{\begin{matrix} 
j & j & J\\
m_2 & m_1 & M 
\end{matrix}\Big\}$ denote the quantum version of Clebsch-Gordan coefficients (q-CG)\cite{kir}. Note that the summation $J \in j \otimes j$ refers to the range  $\{0,1,\ldots 2j\}$.  

The natural challenge is to deduce spectral parameter dependent eigenvalues $\lambda_J(j,j;u)$ for any spin $j$ such that the above relation gives the  known $R^{j,j}(u)$-matrix elements for  $6$-vertex, $19$-vertex and $44$-vertex models. Such a $\lambda_J(j,j;u)$ has been conjectured in Ref.\cite{kaul1}:
\begin{equation}
 \lambda_{J}(j,j;u)=\prod\limits_{k_1=1}^{J}\sinh(k_1\mu-u)\prod\limits_{k_2=J+1}^{2j}\sinh(k_2\mu+u)~,
 \label{eig}
 \end{equation}
 resulting in the spectral parameter dependent  $(R^{j,j})_{m_1,m_2}^{n_1,n_2}(u)$-matrices associated with new vertex models:
 \begin{equation}
(R^{j,j})_{m_1 m_2}^{n_1,n_2}(u)=\sum\limits_{J,M} \Big\{
\begin{matrix} 
j & j & J\\
m_2 & m_1 & M 
\end{matrix}\Big\}\lambda_J(j,j;u)\Big\{
\begin{matrix} 
j & j & J\\
n_1 & n_2 & M 
\end{matrix}\Big\}.
\label{rmat2}
\end{equation}
Here the $SU(2)$ spin $J\in j\otimes j \equiv \{0,1,2,\ldots 2j\}$(allowed  irreducible representations in the tensor product). We have checked, for some values of spin $j$,  that these $R$-matrices obtained from the conjectured form eqn.(\ref{eig}) do obey Yang-Baxter equation and hence are valid Boltzmann weights for new vertex models.
 
So far, we have discussed knot and link invariant computations from  vertex models with edges carrying states of same spin j. We have also seen that there is a neat relation between $R$-matrices with spectral parameter dependent $\lambda_J(j,j;u)$.
Interestingly, the conjectured eigenvalue in eqn.(\ref{eig}) can be generalised to $\lambda_J(j_1,j_2;u)$ where $J\in j_1 \otimes j_2 \equiv \{|j_1-j_2,|j_1-j_2|+1,\ldots j_1+j_2\}$ which will lead to vertex models with adjacent edges carrying states of different spins $j_1\neq j_2$. We refer to these vertex models as bi-partite vertex models(see fig.~\ref{Rm}(b) ). In the following section,  we briefly review bi-partite vertex model and  propose a new algebraic expression for multi-colored link invariants from the associated Boltzmann weights.

\section{Bi-partite vertex model}\label{bipartv}
Let us discuss new vertex model having different spins at the adjacent edges of a lattice which we refer to as `bi-partite vertex model.'

 Following the eigenvalue (eqn.(\ref{eig})) for same spins, the generalisation  $\lambda_J(j_1,j_2;u)$ \cite{kaul1} is
\begin{equation}
 \lambda_{J}(j_1,j_2;u)=\prod\limits_{k_1=|j_1-j_2|+1}^{J}\sinh(k_1\mu-u)\prod\limits_{k_2=J+1}^{j_1+j_2}\sinh(k_2\mu+u)~,
 \label{eig1}
\end{equation}
where $J\in j_1 \times j_2$ and the corresponding spectral parameter dependent $R$-matrices 
(similar to eqn.(\ref{rmat2})) becomes: 
 \begin{equation}
(R^{j_1,j_2})_{m_1 m_2}^{n_1,n_2}(u)=\sum\limits_{J,M} 
\Big\{\begin{matrix} 
j_2 & j_1 & J\\
m_2 & m_1 & M 
\end{matrix}\Big\}\lambda_J(j_1,j_2;u)\Big\{\begin{matrix} 
j_1 & j_2 & J\\
n_1 & n_2 & M 
\end{matrix}\Big\}~.
\label{rmat3}
\end{equation}
The above spectral parameter dependent $R$-matrix must satisfy the following Yang-Baxter equation\cite{yang,baxter} 
\begin{eqnarray}
\sum\limits_{m'_1,m'_2,m'_3}& &(R^{j_1,j_2})_{m_1,m_2}^{m_1',m_2'}(u)(R^{j_1,j_3})_{m'_1,m_3}^{m_1'',m_3'}(u+v)(R^{j_2,j_3})_{m_2',m_3'}^{m_2'',m_3''}(v)\nonumber\\ &=&\sum\limits_{m'_1,m'_2,m'_3}(R^{j_2,j_3})_{m_2,m_3}^{m_2',m_3'}(v)(R^{j_1,j_3})_{m_1,m_3'}^{m_1',m_3''}(u+v)(R^{j_1,j_2})_{m_1',m_2'}^{m_1'',m_2''}(u)~.
\label{ybe1}
\end{eqnarray}
We have checked for some values of $j_1,j_2,j_3$ values that the conjectured form of $R$-matrices (\ref{rmat3})  indeed obey the above Yang-Baxter equation. 

Taking the limit
$u,v,u+v \rightarrow \infty$ on $(R^{j_1,j_2})_{m_1,m_2}^{m_1',m_2'}(u)$ and a suitable normalisation 
$\mathcal N=(R^{j_1,j_2})_{j_1,j_2}^{j_1,j_2}(u\rightarrow \infty)$, we obtain spectral parameter independent matrix elements.
Multiplying an appropriate permutation matrix $\hat P^{j_1,j_2}$, the matrix
\begin{equation}
(\hat R^{j_1,j_2})_{m_1,m_2}^{n_1 n_2}= {1 \over \mathcal N} (\hat P^{j_1,j_2})_{m_1,m_2}^{m_1', m_2'} (R^{j_1,j_2})_{m_1',m_2'}^{n_1,n_2}(u \rightarrow \infty),
\end{equation}
define braiding generators $b(j_1,j_2)$ whose action on two-strands with representations $j_1,j_2$ will be 
\begin{equation}
b(j_1,j_2)\vert j_1,j_2\rangle \propto \vert j_2,j_1\rangle~.
\end{equation}
Arbitrary braid word using these generators must keep track of the spin $j_1,j_2,\ldots j_n$ on the $n$-strands.The collection of such braid words actually forms a groupoid\cite{kaul1994}. Further, closure will require the initial state $|j_1,j_2\ldots j_n\rangle$ to be same as the final state after the operation of braid word. Such a closure of braid word will result in multi-component links carrying different representations. Using the matrix form of the braiding generators 
$b(j_1,j_2), b(j_1,j_3)\ldots$, derived from bi-partite vertex models, we can obtain multi-colored link invariants for component knots  carrying different representations. We illustrate this procedure for simple links by explicitly writing down the Boltzmann weights $R^{j_1=1,j_2=1/2}(u)$ in the following section.
\subsection{R-matrix for different spin}
For the calculation of multi-component link invariant it is essential to determine  $(R^{j_1,j_2})_{m_1,m_2}^{n_1,n_2}$ matix for different $j_1$ and $j_2$.
As an example, let us take $j_1=1$ and $j_2=1/2$ where the spectral parameter dependent eigenvalues (\ref{eig1}) are
$$\lambda_{1/2}(u) =sinh({3 \mu\over 2} +u)  ~{\rm and} ~ \lambda_{3/2}(u)=sinh({3\mu \over 2} - u)~.$$
Using these eigenvalues, we obtain the following  $R^{1,1/2}(u)$-matrix (\ref{rmat3}):
\begin{equation}
(R^{1,\frac{1}{2}})_{m_1,m_2}^{n_1,n_2}(u)= \left(
\begin{array}{c|ccccccc}
 m_1,m_2\backslash n_1,n_2\rightarrow        & 1,\frac{1}{2} & 1,\frac{-1}{2} & 0,\frac{1}{2} & 0,\frac{-1}{2} & -1,\frac{1}{2} & -1,\frac{-1}{2}  \\
 \hline\\
1,\frac{1}{2} & x_1(u) & 0 & 0 & 0 & 0 & 0 \\
 1,\frac{-1}{2} &0 & x_2(u) & x_3'(u) & 0 & 0 & 0 \\
 0,\frac{1}{2} & 0 & x_3(u) & x'_2(u) & 0 & 0 & 0 \\
 0,\frac{-1}{2} &0 & 0 & 0 & x_2'(u) & x_3'(u) & 0 \\
-1,\frac{1}{2} & 0 & 0 & 0 & x_3(u) & x_2(u) & 0 \\
-1,\frac{-1}{2} & 0 & 0 & 0 & 0 & 0 & x_1(u) \\
\end{array}
\right)~,
\end{equation}
where
\begin{eqnarray*}
x_1(u)&=&\sinh(\frac{3\mu}{2}-u) ,x_2(u)=-\sinh(\frac{\mu}{2}+u),~ x_3(u)=(\sinh{2\mu}\sinh{\mu})^{\frac{1}{2}}e^{-u},\\
x'_3(u)&=&(\sinh{2\mu}\sinh{\mu})^{\frac{1}{2}}e^{u},  x'_2(u)=\sinh(\frac{\mu}{2}-u)~.
\end{eqnarray*}
Substituting the limit as $u\rightarrow\infty$ and $q=e^{2\mu}$ for $j_1=1$, $j_2=\frac{1}{2}$, we get
\begin{eqnarray}
\lim_{u\rightarrow \infty }\frac{(R^{1,\frac{1}{2}})_{m_1,m_2}^{n_1,n_2}(u)}{(R^{1,\frac{1}{2}})_{1,\frac{1}{2}}^{1,\frac{1}{2}}(u)}=~~~~~~~~~~~~~~~~~~~~~~~~~~~~~~~~~~~~~~~~~~~~~~~~~~~~~~~~~~~~~~~~~~\nonumber\\
 \left(
\begin{array}{c|ccccccc}
 m_1,m_2\backslash n_1,n_2\rightarrow        & 1,\frac{1}{2} & 1,\frac{-1}{2} & 0,\frac{1}{2} & 0,\frac{-1}{2} & -1,\frac{1}{2} & -1,\frac{-1}{2}  \\
 \hline\\
1,\frac{1}{2} & 1 & 0 & 0 & 0 & 0 & 0 \\
 1,\frac{-1}{2} &0 & q & (1-q) \sqrt{1+q} & 0 & 0 & 0 \\
 0,\frac{1}{2} & 0 & 0 & \sqrt{q} & 0 & 0 & 0 \\
 0,\frac{-1}{2} &0 & 0 & 0 & \sqrt{q} & (1-q) \sqrt{1+q} & 0 \\
-1,\frac{1}{2} & 0 & 0 & 0 & 0 & q & 0 \\
-1,\frac{-1}{2} & 0 & 0 & 0 & 0 & 0 & 1 \\
\end{array}
\right).
\end{eqnarray}
In order to obtain braiding generators $b(j_1=1,j_2=1/2)$, we need a suitable permutation matrix $\hat P^{j_1=1,j_2=1/2}$ so that the sequence of states mentioned along the row and column in the above $R^{j_1,j_2}$-matrix are maintained. This leads to the following proposition.

{\bf Proposition 1}:The permutation matrix $\hat P^{j_1 j_2}$ action on the column state 
 \begin{equation}
 \hat P^{j_1,j_2}
 \begin{bmatrix} 
    |j_1,j_2\rangle  \\
    |j_1,j_2-1\rangle  \\
    \vdots  \\
    |j_1,-j_2\rangle  \\
    |j_1-1,j_2\rangle  \\
    |j_1-1,j_2-1\rangle  \\
     \vdots  \\
    |-j_1,-j_2\rangle  \\ 
    \end{bmatrix}=
    \begin{bmatrix} 
    |j_1,j_2\rangle  \\
    |j_1-1,j_2\rangle  \\
    \vdots  \\
    |-j_1,j_2\rangle  \\
    |j_1,j_2-1\rangle  \\
    |j_1-1,j_2-1\rangle  \\
     \vdots  \\
    |-j_1,-j_2\rangle  \\ 
    \end{bmatrix}.
    \label{permu}
 \end{equation}
For $j_1=1.j_2=1/2$, the $\hat P^{1,1/2}$ will be
\begin{equation}
\hat P^{1,\frac{1}{2}}= \left(
\begin{array}{ccccccc}
 1 & 0 & 0 & 0 & 0 & 0 \\
 0 & 0 & 1 & 0 & 0 & 0 \\
 0 & 0 & 0 & 0 & 1 & 0 \\
 0 & 1 & 0 & 0 & 0 & 0 \\
 0 & 0 & 0 & 1 & 0 & 0 \\
 0 & 0 & 0 & 0 & 0 & 1 \\
\end{array}
\right).
\end{equation}
We will now use the following permutation matrix  in the braiding generator construction:
\begin{equation}
(\hat{R}^{j_1,j_2})_{m_1,m_2}^{n_1,n_2}= ({\hat P}^{j_1,j_2})_{m_1,m_2}^{m_1',m_2'} \lim_{u \rightarrow \infty} \frac{(R^{j_1,j_2})_{m_1',m_2'}^{n_1,n_2}(u)}{(R^{j_1,j_2})_{j_1,j_2}^{j_1,j_2}(u)}.
\label{hat1}
\end{equation}
The explicit  form of 
$\hat{R}^{1,\frac{1}{2}}$ matrix is
\begin{equation}
(\hat R^{1,\frac{1}{2})_{m_1,m_2}^{n_1,n_2}}= \left(
\begin{array}{ccccccc}
  1 & 0 & 0 & 0 & 0 & 0 \\
 0 & 0 & \sqrt{q} & 0 & 0 & 0 \\
 0 & 0 & 0 & 0 & q & 0 \\
 0 & q & (1-q) \sqrt{1+q} & 0 & 0 & 0 \\
 0 & 0 & 0 & \sqrt{q} & (1-q) \sqrt{1+q} & 0 \\
 0 & 0 & 0 & 0 & 0 & 1 \\
\end{array}
\right).
\end{equation}
Similar construction of $\hat{R}^{{1 \over 2},1}$ matrix for $j_1=1/2,j_2=1$ turns out 
to be  transpose of matrix $\hat{R}^{1,{1 \over 2}}$. Note that the identity matrix
can be written as:
\begin{equation}
\hat R^{j_1,j_2}. \left[\hat R^{j_1,j_2}\right]^{-1}=\hat R^{j_1,j_2}. \left[[\hat R^{j_2,j_1}]^{\intercal}\right]^{-1}=\mathbb I~.
\end{equation}
Hence we can write the matrix representation of the braiding generators $b_i[j_1,j_2]$ of the groupoid obeying
$$b_i[j_1,j_2]\left(b_i[j_2,j_1]\right)^{-1}=\mathbb I~,$$ as follows:
\begin{eqnarray}
b_i[j_1,j_2]&=&\underbrace{ \mathbb I_1 \times \mathbb I_2 \times \ldots \mathbb I\times}_{i-1}
 \hat{R}^{j_1,j_2}\times \mathbb I_{i+2} \ldots_ \\
b_i[j_1,j_2] ^{-1}&=&\underbrace{ \mathbb I_1 \times \mathbb I_2 \times \ldots \mathbb I\times}_{i-1}  (\hat{R}^{j_2,j_1})^{-1}\times \mathbb I_{i+2} \ldots. \label{condn}
\end{eqnarray}
Hence for any braid word $A$, whose closure will give multi-component links, we will use the above matrix representation for braiding generators and their inverses. 
Similar to the knot invariants (eqn.(\ref{formu4})), we propose the following formulae for multi-colored link invariants where the component knots carry different spins.

{\bf Proposition 2:} The multi-colored link invariants ${\tilde \alpha}_{j_1,j_2,\ldots j_n}(A)$(upto an overall factor of power of $q^{1/2}$) for any $n$-component link ${\mathcal L}$ with different spins, obtained from closure of any $r$-strand braid word $A$ is given by
\begin{equation}
\alpha_{j_1,j_2,\ldots j_n}{[A({\mathcal L})]}=q^{{1\over 2}\mathcal C} {\tilde \alpha}_{j_1,j_2,\ldots j_n}{(A)}=q^{{1\over 2}\mathcal C} \prod_{i=1}^n(\tau_{j_i}\bar {\tau}_{j_i})^{-\ell_i/2} {\rm Tr}\{H.A \}~,
\label{formu1}
\end{equation}
where the first factor gives an overall $q$-dependent normalisation with integer $\mathcal C$  dependent on the spins, writhe of the component knots and the linking number between component knots of a link. The $\ell_i$'s are the number of times spin $j_i$ occurs in the $r$-strand braid $A$. That is, $\sum_{i=1}^n \ell_i =r$. Further, the matrix form of $H$ will depend on the order of such repeated spins occurring in the $r$-strand braid. For instance, a 3-strand braid with spin $j_1$ on first strand, $j_2\neq j_1$ on second strand and again $j_1$ on first strand will mean 
$$H= h_{j_1} \otimes h_{j_2}\otimes h_{j_1}~.$$
We must again follow sequence of  matrix operations for  braid word $A$ similar to the sequence (\ref{ordbraid}) explained for same spins. Recall the definitions of $h_{j_i}$'s (eqn.(\ref{hmat})) 
$\tau_{j_i}$'s and ${\bar {\tau}}_{j_i}$'s(eqn.(\ref{tau})) as discussed in section \ref{vertexm}. We will explicitly work out multi-colored link invariants for some links in the following subsection.
\subsubsection{Multi-colored link invariants}
For two component links, with spin $j_1=1$ on first component and spin ${1/2}$ on second component, we need to write the $n$-strand braid word keeping track of the spins. 
\begin{enumerate}
\item For the simplest Hopf link ${\mathcal H}$ obtained from closure of  two-strand braid, the matrix operation will be 
$$A({\mathcal H})=b_1[1/2,1].b_1[1,1/2]={\hat R}^{1/2,1}{\hat R}^{1,1/2}$$
and $H=h_1 \otimes h_{1/2}$ giving 
\begin{equation}
 \tilde {\alpha}_{1,1/2}[A({\mathcal H})]= \frac{1 + q + q^2 + q^3 + q^4 + q^5}{q^{3/2}}
\end{equation}
which agrees with multi-colored  Jones polynomial computed from $SU(2)$ Chern-Simons theory upto overall factor.
\item The other familiar two component link is the Whitehead Link $W$ obtained from closure of three-strand braid. Using the following matrix operation for $A(W)= b_ 2^{-1} (1, 1/2).b_ 2^{-1} (1/2, 1).b_ 1 (1, 1/2). b_ 2^{-1} (1/2, 1/2) .b_ 1 (1/2, 1)$ and $H=h_{1/2} \otimes h_1
\otimes h_{1/2}$ giving
\begin{equation}
{\tilde \alpha}_{1/2,1}[A(W)]=-\frac{-1+q^2+q^3+2 q^4+q^5+q^6+q^9}{q^5}~.
\end{equation}
\item See Table.\ref{diff} where we have presented the braid word and the multi-colored invariant for link $L7a3$ whose results are matching with $SU(N=2)$ results in Ref. \cite{dhara2018multi} upto a overall factor 
\end{enumerate}
We have also worked out ${\tilde \alpha}_{j_1,j_2,j_3}$ for Borrowmean rings when 
$j_1=j_2\neq j_3$ and $j_1\neq j_2\neq j_3$ for $j_1=1/2,j_2=1,j_3=3/2$. As this computation requires $\hat R^{j_1,j_2}$ for $j_1=1,j_2=3/2$ and $j_1=1/2,j_2=3/2$, we have presented the matrix elements in Appendix A. We have tabulated the explicit multi-colored link invariants for 
some $j_1,j_2,j_3$ in Table. \ref{diff}.\\
\vspace{1cm}
{\large\begin{table}[ht]
\begin{center}
\begin{tabular}{|c|c|c|c|}
\hline
Link       & Braidword & $j_1,j_2$ & Polynomial  \\
\hline
{}       & {}  & $1,\frac{1}{2}$& $q^{-\frac{3}{2}}\left(1 + q + q^2 + q^3 + q^4 + q^5\right)$  \\
&&&\\
Hopf Link & $b_1.b_1$  & $1,\frac{3}{2}$& $q^{-\frac{5}{2}}\Big((1 + q + q^2) (1 + q + q^2 + q^3) $  \\
{}&{}&{}&$\times (1 - q + q^3 - q^5 + q^6)\Big)$\\
&&&\\
{}      & {}  & $\frac{3}{2},\frac{1}{2}$& $q^{-2}(1 + q) (1 + q^2) (1 + q^4)$ \\
\hline
\end{tabular}
\end{center}
\label{diff}
\end{table}}

\begin{table}
\begin{center}
\begin{tabular}{|c|c|c|c|}
\hline
Link       & Braidword & $j_1,j_2,j_3$ & Polynomial  \\
\hline
{}      & {}  & $1,\frac{1}{2},\frac{1}{2}$ & $q^{-4}\left(1 + q^2 + q^3 + q^4 + 2 q^5 + q^7 - q^{11}\right)$ \\
&&&\\
L7a3 Link & $b_1.b_2^{-1}.b_1^3.b_2^{-1}.b_1$  & $\frac{1}{2},1,1$ & $q^{-\frac{9}{2}}\big(1 + 2 q^3 + 2 q^4 + q^5 + 2 q^6 + q^7$ \\
 {}&{}&{}&$ - q^8 - q^{10} - q^{12} + q^{14} - 
 2 q^{15} + q^{17}\big)$\\
 &&&\\
 \hline
{}    & {} & $\frac{1}{2},1,\frac{1}{2}$& $q^{-5}\left(-1 + q^2 + q^3 + 2 q^4 + q^5 + q^6 + q^9\right)$\\
&&&\\
Whitehead  link    & $b_1.b_2^{-1}.b_1.b_2^{-2}$ & $1,\frac{3}{2},1$& $q^{-\frac{25}{2}}(1 + q) (1 - q - q^2 + q^7 + q^8 + q^{10} $\\
{}&{}&{}&$+ q^{11} + q^{13} + q^{14} + q^{17} -
    q^{20} + q^{21})$\\
    &&&\\
\hline
   & {} & $\frac{1}{2},1,\frac{1}{2}$ &$ 2-\frac{1}{q^5}+\frac{1}{q^4}+\frac{1}{q^3}+\frac{1}{q^2}+\frac{3}{q}$ \\
 {}&{}&{}&$+3 q+q^2+q^3+q^4-q^5$\\
 &&&\\
  Borromean Ring    & $b_1^{-1}.b_2.b_1^{-1}.b_2.b_1^{-1}.b_2$ &$\frac{1}{2},1,\frac{3}{2}$& $-q^{-8}(1 + q + q^2) (1 - q - 2 q^4 - q^6 $ \\
   {}&{}&{}&$- 2 q^7 - q^8 - 2 q^{10} - q^{13} + 
   q^{14})$\\
   &&&\\
 \hline
\end{tabular}
\end{center}
\caption{ }\label{diff}
\end{table}
\newpage
\section{Conclusion}
In this paper, we have shown efficient computation of multi-colored link invariants from the braid group representations derived from new bi-partite vertex models. Here the adjacent edges carry different spins as shown in fig.~\ref{Rm}(b).  These invariants are proportional to multi-colored Jones' polynomials in the literature.  

Instead of $SU(2)$ group involving spin $j_1,j_2 $ states on the edges intersecting the vertex, we could place states of $SU(N)$ representations. The procedure presented in the paper must be generalisable for  SU(N) group resulting in new vertex models and their link invariants.
We hope to report in future on such types of vertex models and the link invariant computations.
These invariants are known in the literature as multi-colored HOMFLY-PT polynomials. 

{\bf Acknowledgements}
SD would like to thank CSIR for a research fellowship. RKK thanks Department of Atomic Energy, Government of India for financial support. PR would like to thank  Kavli Institute for Theoretical Physics at the University of California  Santa Barbara for the wonderful research environment and hospitality where this manuscript was completed. This research was supported in part by the National Science
Foundation under Grant No. PHY-1748958.
\vskip1cm

\appendix{\bf Appendix A}\\

We present the ${\hat R^{1,3/2}, \hat R^{1/2,3/2}}$ which will be useful for multi-colored link invariants for other representations. From eqn.(\ref{rmat3}), the $R$-matrix when $u\rightarrow \infty$ is
\begin{eqnarray*}
\lim_{u\rightarrow \infty} \frac{(R^{1,\frac{1}{2}})_{m_1,m_2}^{n_1,n_2}(u)}{(R^{1,\frac{1}{2}})_{1,\frac{1}{2}}^{1,\frac{1}{2}}(u)}= ~~~~~~~~~~~~~~~~~~~~~~~~~~~~~~~~~~~~~~~~~~~~~~~~~~~~~~~~~~~~~~~~~~~\\
\left(
\begin{array}{c|ccccccc}
 m_1,m_2\backslash n_1,n_2\rightarrow        & 1,\frac{1}{2} & 1,\frac{-1}{2} & 0,\frac{1}{2} & 0,\frac{-1}{2} & -1,\frac{1}{2} & -1,\frac{-1}{2}  \\
 \hline\\
1,\frac{1}{2} & 1 & 0 & 0 & 0 & 0 & 0 \\
 1,\frac{-1}{2} &0 & q & (1-q) \sqrt{1+q} & 0 & 0 & 0 \\
 0,\frac{1}{2} & 0 & 0 & \sqrt{q} & 0 & 0 & 0 \\
 0,\frac{-1}{2} &0 & 0 & 0 & \sqrt{q} & (1-q) \sqrt{1+q} & 0 \\
-1,\frac{1}{2} & 0 & 0 & 0 & 0 & q & 0 \\
-1,\frac{-1}{2} & 0 & 0 & 0 & 0 & 0 & 1 \\
\end{array}
\right).
\end{eqnarray*}
\begin{eqnarray*}
\lim_{u\rightarrow \infty} \frac{(R^{\frac{3}{2},\frac{1}{2}})_{m_1,m_2}^{n_1,n_2}(u)}{(R^{\frac{3}{2},\frac{1}{2}})_{\frac{3}{2},\frac{1}{2}}^{\frac{3}{2},\frac{1}{2}}(u)}=~~~~~~~~~~~~~~~~~~~~~~~~~~~~~~~~~~~~~~~~~~~~~~~~~~~~~~~~~~~~~~~~~~~~~~~~~~\\
 \left(
\begin{array}{c|ccccccccc}
 m_1,m_2\backslash n_1,n_2\rightarrow        & \frac{3}{2},\frac{1}{2} & \frac{3}{2},\frac{-1}{2} & \frac{1}{2},\frac{1}{2} & \frac{1}{2},\frac{-1}{2} & \frac{-1}{2},\frac{1}{2} & \frac{-1}{2},\frac{-1}{2} & \frac{-3}{2},\frac{1}{2} & \frac{-3}{2},\frac{-1}{2}  \\
 \hline\\
 \frac{3}{2},\frac{1}{2} & 1 & 0 & 0 & 0 & 0 & 0 & 0 & 0 \\
 \frac{3}{2},\frac{-1}{2} &0 & y_1 & y_2 & 0 & 0 & 0 & 0 & 0 \\
 \frac{1}{2},\frac{1}{2} &0 & 0 & y_3 & 0 & 0 & 0 & 0 & 0 \\
\frac{1}{2},\frac{-1}{2} & 0 & 0 & 0 & y_5 & y_4 & 0 & 0 & 0 \\
 \frac{-1}{2},\frac{1}{2} &0 & 0 & 0 & 0 & y_5 & 0 & 0 & 0 \\
 \frac{-1}{2},\frac{-1}{2} &0 & 0 & 0 & 0 & 0 & y_3 & y_2 & 0 \\
 \frac{-3}{2},\frac{1}{2} &0 & 0 & 0 & 0 & 0 & 0 & y_1 & 0 \\
 \frac{-3}{2},\frac{-1}{2} &0 & 0 & 0 & 0 & 0 & 0 & 0 & 1 \\
\end{array}
\right).
\end{eqnarray*}
 where $y_1=q^{3/2}$, $y_2=(1-q) \sqrt{1+q+q^2}$, $y_3=\sqrt{q}$ ,$y_4=1-q^2$, $y_5=q$.\\
 
 Similarly  
$\lim_{u\rightarrow \infty}\frac{(R^{\frac{3}{2},1})_{m_1,m_2}^{n_1,n_2}(u)}{(R^{\frac{3}{2},1})_{\frac{3}{2},1}^{\frac{3}{2},1}(u)}=$
\begin{equation}
 \left(
\begin{array}{c|ccccccccccccc}
 m_1,m_2\backslash n_1,n_2\rightarrow        & \frac{3}{2},1 & \frac{3}{2},0 & \frac{3}{2},-1 &\frac{1}{2},1 & \frac{1}{2},0 & \frac{1}{2},-1 &\frac{-1}{2},1 & \frac{-1}{2},0 & \frac{-1}{2},-1 & \frac{-3}{2},1 & \frac{-3}{2},0 & \frac{-3}{2},-1 & \\
 \hline\\
 \frac{3}{2},1 &1 & 0 & 0 & 0 & 0 & 0 & 0 & 0 & 0 & 0 & 0 & 0 \\
 \frac{3}{2},0 &0 & z_1 & 0 &  z_2 & 0 & 0 & 0 & 0 & 0 & 0 & 0 & 0 \\
 \frac{3}{2},-1 &0 & 0 & z_6 & 0 & z_3 & 0 & z_4 & 0 & 0 & 0 & 0 & 0 \\
 \frac{1}{2},1 &0 & 0 & 0 & z_8 & 0 & 0 & 0 & 0 & 0 & 0 & 0 & 0 \\
 \frac{1}{2},0 &0 & 0 & 0 & 0 & z_1 & 0 & z_5 & 0 & 0 & 0 & 0 & 0 \\
 \frac{1}{2},-1 &0 & 0 & 0 & 0 & 0 & z_7 & 0 & z_5 & 0 & z_4 & 0 & 0 \\
 \frac{-1}{2},1 &0 & 0 & 0 & 0 & 0 & 0 & z_7 & 0 & 0 & 0 & 0 & 0 \\
 \frac{-1}{2},0 &0 & 0 & 0 & 0 & 0 & 0 & 0 & z_1 & 0 & z_3 & 0 & 0 \\
 \frac{-1}{2},-1 &0 & 0 & 0 & 0 & 0 & 0 & 0 & 0 & z_8 & 0 & z_2 & 0 \\
 \frac{-3}{2},1 &0 & 0 & 0 & 0 & 0 & 0 & 0 & 0 & 0 & z_6 & 0 & 0 \\
 \frac{-3}{2},0 &0 & 0 & 0 & 0 & 0 & 0 & 0 & 0 & 0 & 0 & z_1 & 0 \\
 \frac{-3}{2},-1 &0 & 0 & 0 & 0 & 0 & 0 & 0 & 0 & 0 & 0 & 0 & 1 \\
\end{array}
\right).
\end{equation}
where
\begin{eqnarray*}
z_1&=&q^{3/2}, z_2=(1-q) \sqrt{(1+q) \left(1+q+q^2\right)}, z_3=(1-q) q \sqrt{(1+q) \left(1+q+q^2\right)},\\
 z_4&=&(-1+q)^2 (1+q) \sqrt{1+q+q^2}, z_5=(1-q) \sqrt{q} (1+q)^{3/2}, z_6=q^3, z_7=q^2, 
 z_8=q~.\\
\end{eqnarray*}
Using the proposition 1 (\ref{permu}), the  permutation matrices are 
\begin{equation*}
\hat P^{1,\frac{1}{2}}= \left(
\begin{array}{ccccccc}
 1 & 0 & 0 & 0 & 0 & 0 \\
 0 & 0 & 1 & 0 & 0 & 0 \\
 0 & 0 & 0 & 0 & 1 & 0 \\
 0 & 1 & 0 & 0 & 0 & 0 \\
 0 & 0 & 0 & 1 & 0 & 0 \\
 0 & 0 & 0 & 0 & 0 & 1 \\
\end{array}
\right),~
\hat P^{\frac{3}{2},\frac{1}{2}}= \left(
\begin{array}{cccccccc}
 1 & 0 & 0 & 0 & 0 & 0 & 0 & 0 \\
 0 & 0 & 1 & 0 & 0 & 0 & 0 & 0 \\
 0 & 0 & 0 & 0 & 1 & 0 & 0 & 0 \\
 0 & 0 & 0 & 0 & 0 & 0 & 1 & 0 \\
 0 & 1 & 0 & 0 & 0 & 0 & 0 & 0 \\
 0 & 0 & 0 & 1 & 0 & 0 & 0 & 0 \\
 0 & 0 & 0 & 0 & 0 & 1 & 0 & 0 \\
 0 & 0 & 0 & 0 & 0 & 0 & 0 & 1
\end{array}
\right),
~
\hat P^{\frac{3}{2},1}= \left(
\begin{array}{ccccccccccccc}
 1 & 0 & 0 & 0 & 0 & 0 & 0 & 0 & 0 & 0 & 0 & 0 \\
 0 & 0 & 0 & 1 & 0 & 0 & 0 & 0 & 0 & 0 & 0 & 0 \\
 0 & 0 & 0 & 0 & 0 & 0 & 1 & 0 & 0 & 0 & 0 & 0 \\
 0 & 0 & 0 & 0 & 0 & 0 & 0 & 0 & 0 & 1 & 0 & 0 \\
 0 & 1 & 0 & 0 & 0 & 0 & 0 & 0 & 0 & 0 & 0 & 0 \\
 0 & 0 & 0 & 0 & 1 & 0 & 0 & 0 & 0 & 0 & 0 & 0 \\
 0 & 0 & 0 & 0 & 0 & 0 & 1 & 1 & 0 & 0 & 0 & 0 \\
 0 & 0 & 0 & 0 & 0 & 0 & 0 & 0 & 0 & 0 & 0 & 0 \\
 0 & 0 & 1 & 0 & 0 & 0 & 0 & 0 & 0 & 0 & 1 & 0 \\
 0 & 0 & 0 & 0 & 0 & 1 & 0 & 0 & 0 & 0 & 0 & 0 \\
 0 & 0 & 0 & 0 & 0 & 0 & 0 & 0 & 1 & 0 & 0 & 0 \\
 0 & 0 & 0 & 0 & 0 & 0 & 0 & 0 & 0 & 0 & 0 & 1
\end{array}
\right).
\end{equation*}
The $\hat{R}^{j_1,j_2}$matrix using eqn.(\ref{hat1}) is given by;
\begin{equation*}
(\hat R^{1,\frac{1}{2}})_{m_1,m_2}^{n_1,n_2}= \left(
\begin{array}{ccccccc}
  1 & 0 & 0 & 0 & 0 & 0 \\
 0 & 0 & \sqrt{q} & 0 & 0 & 0 \\
 0 & 0 & 0 & 0 & q & 0 \\
 0 & q & (1-q) \sqrt{1+q} & 0 & 0 & 0 \\
 0 & 0 & 0 & \sqrt{q} & (1-q) \sqrt{1+q} & 0 \\
 0 & 0 & 0 & 0 & 0 & 1 \\
\end{array}
\right),
\end{equation*}

\begin{equation*}
(\hat R^{\frac{3}{2},\frac{1}{2}})_{m_1,m_2}^{n_1,n_2}= \left(
\begin{array}{cccccccc}
 1 & 0 & 0 & 0 & 0 & 0 & 0 & 0 \\
 0 & 0 & \sqrt{q} & 0 & 0 & 0 & 0 & 0 \\
 0 & 0 & 0 & 0 & q & 0 & 0 & 0 \\
 0 & 0 & 0 & 0 & 0 & 0 & q^{3/2} & 0 \\
 0 & q^{3/2} & (1-q) \sqrt{1+q+q^2} & 0 & 0 & 0 & 0 & 0 \\
 0 & 0 & 0 & q & 1-q^2 & 0 & 0 & 0 \\
 0 & 0 & 0 & 0 & 0 & \sqrt{q} & (1-q) \sqrt{1+q+q^2} & 0 \\
 0 & 0 & 0 & 0 & 0 & 0 & 0 & 1 \\
\end{array}
\right).
\end{equation*}
\begin{equation*}
(\hat R^{\frac{3}{2},\frac{1}{2}})_{m_1,m_2}^{n_1,n_2}= \left(
\begin{array}{ccccccccccccc}
 1 & 0 & 0 & 0 & 0 & 0 & 0 & 0 & 0 & 0 & 0 & 0 \\
 0 & 0 & 0 & z_8 & 0 & 0 & 0 & 0 & 0 & 0 & 0 & 0 \\
 0 & 0 & 0 & 0 & 0 & 0 & z_7 & 0 & 0 & 0 & 0 & 0 \\
 0 & 0 & 0 & 0 & 0 & 0 & 0 & 0 & 0 & z_6 & 0 & 0 \\
 0 & z_1 & 0 & z_2 & 0 & 0 & 0 & 0 & 0 & 0 & 0 & 0 \\
 0 & 0 & 0 & 0 & z_1 & 0 & z_5 & 0 & 0 & 0 & 0 & 0 \\
 0 & 0 & 0 & 0 & 0 & 0 & 0 & z_1 & 0 & z_3 & 0 & 0 \\
 0 & 0 & 0 & 0 & 0 & 0 & 0 & 0 & 0 & 0 & z_1 & 0 \\
 0 & 0 & z_6 & 0 & z_3 & 0 & z_4 & 0 & 0 & 0 & 0 & 0 \\
 0 & 0 & 0 & 0 & 0 & z_7 & 0 & z_5 & 0 & z_4 & 0 & 0 \\
 0 & 0 & 0 & 0 & 0 & 0 & 0 & 0 & z_8 & 0 & z_2 & 0 \\
 0 & 0 & 0 & 0 & 0 & 0 & 0 & 0 & 0 & 0 & 0 & 1 \\

\end{array}
\right).
\end{equation*}
Using these $\hat R$-matrices, the multi-colored link invariants can be efficiently computed.
\bibliography{main}
\bibliographystyle{unsrt}
\end{document}